\documentclass[final,nofootinbib,aps,twocolumn,showpacs,showkeys,groupaddress,preprintnumbers,floatfix]{revtex4-1}

\usepackage{dynlearn}

\newcommand{\source} [1] {\ensuremath{X_{#1}}\xspace}
\newcommand{\target}     {\ensuremath{Y}\xspace}

\newcommand{\sep}{\cdot}

\newcommand{\PID} [3] {
  \ensuremath{
    \operatorname{I_{#1}}
    \if\relax\detokenize{#3}\relax
      \!\!
    \else
      \left[ #2 \rightarrow #3 \right]
    \fi
  }
  \xspace
}

\newcommand{\Ipart}  [2] [\target] {\PID{\partial}{#2}{#1}}

\newcommand{\Ibroja} [2] [\target] {\PID{\textrm{BROJA}}{#2}{#1}}

\newcommand{\pntunqdist}{
  \begin{tabular}[t]{cccc}
    \multicolumn{4}{c}{\textsc{Pnt. Unq.}}                           \\
    \toprule
    $\source{0}$ & $\source{1}$ & $\target$ & $\Pr$             \\
    \midrule
    0            & 1            & 1         & $\nicefrac{1}{4}$ \\
    1            & 0            & 1         & $\nicefrac{1}{4}$ \\
    0            & 2            & 2         & $\nicefrac{1}{4}$ \\
    2            & 0            & 2         & $\nicefrac{1}{4}$ \\
    \bottomrule
  \end{tabular}
}

\newcommand{\SKARzero} [3] {\ensuremath{\operatorname{S}(#1 : #2 ~||~ #3)}\xspace}
\newcommand{\SKARone}  [3] {\ensuremath{\operatorname{S}(#1 \rightarrow #2 ~||~ #3)}\xspace}
\newcommand{\SKARtwo}  [3] {\ensuremath{\operatorname{S}(#1 \leftrightarrow #2 ~||~ #3)}\xspace}
 
\begin{document}

\def\ourTitle{
  A Perspective on Unique Information:\\
	Directionality, Intuitions, and Secret Key Agreement
}

\def\ourAbstract{
Recently, the partial information decomposition emerged as a promising framework for identifying the meaningful components of the information contained in a joint distribution.
Its adoption and practical application, however, have been stymied by the lack of a generally-accepted method of quantifying its components.
Here, we briefly discuss the bivariate (two-source) partial information decomposition and two implicitly directional interpretations used to intuitively motivate alternative component definitions.
Drawing parallels with secret key agreement rates from information-theoretic cryptography, we demonstrate that these intuitions are mutually incompatible and suggest that this underlies the persistence of competing definitions and interpretations.
Having highlighted this hitherto unacknowledged issue, we outline several possible solutions.
}

\def\ourKeywords{
  information theory, partial information decomposition, secret key agreement, cryptography
}

\hypersetup{
  pdfauthor={James P. Crutchfield},
  pdftitle={\ourTitle},
  pdfsubject={\ourAbstract},
  pdfkeywords={\ourKeywords},
  pdfproducer={},
  pdfcreator={}
}

\author{Ryan G. James}
\email{rgjames@ucdavis.edu}

\author{Jeffrey Emenheiser}
\email{jemenheiser@ucdavis.edu}

\author{James P. Crutchfield}
\email{chaos@ucdavis.edu}

\affiliation{Complexity Sciences Center and Physics Department,
University of California at Davis, One Shields Avenue, Davis, CA 95616}

\date{\today}
\bibliographystyle{unsrt}

\title{\ourTitle}

\begin{abstract}
\ourAbstract
\end{abstract}



\preprint{\arxiv{1808.XXXX}}

\title{\ourTitle}
\date{\today}
\maketitle

\setstretch{1.1}


\section{Introduction}
\label{sec:introduction}

Consider a joint distribution over ``source'' variables \source{0} and \source{1} and ``target'' \target.
Such distributions arise in many settings: sensory integration, logical computing, neural coding, functional network inference, and many others.
One promising approach to understanding how the information shared between \source{0}, \source{1}, and \target is organized is the \emph{partial information decomposition} (PID)~\cite{williams2010nonnegative}.
This decomposition seeks to quantify how much of the information shared between \source{0}, \source{1}, and \target is done so \emph{redundantly}, how much is \emph{uniquely} attributable to \source{0}, how much is \emph{uniquely} attributable to \source{1}, and finally how much arises \emph{synergistically} by considering both \source{0} and \source{1} together.

Unfortunately, the lack of a commonly accepted method of quantifying these components has hindered PID's adoption.
In point of fact, several proposed axioms are not mutually consistent.
And, to date, there is little agreement as to which should hold.
Here, we take a step toward rectifying these issues by bringing to light a potentially fundamental inconsistency in the intuitions commonly and often implicitly brought to bear upon information decomposition.
We make the intuitions quantitative by appealing to information-theoretic cryptography.
Taken together, our observations suggest that the context in which PID is applied should determine how its components are quantified.

Our development proceeds as follows.
\Cref{sec:pid} briefly describes the two-source PID.
\Cref{sec:intuitions} calls out the two distinct intuitions often used in interpreting PID.
\Cref{sec:distribution} introduces a prototype distribution that highlights the issues and we interpret it through the lenses of the two intuitions.
\Cref{sec:secrets} defines secret key agreement rates and computes them for the prototype distribution.
\Cref{sec:natural} then discusses how the two intuitions relate to secret key agreement rates and identifies when the latter result in viable decompositions.
Finally, \cref{sec:conclusion} summarizes our findings and speculates as to how future developments can bring consistency to PID.

\section{Partial Information Decomposition}
\label{sec:pid}

Two-source PID seeks to decompose the mutual information $\I{\source{0}\source{1} : \target}$ between ``sources'' \source{0} and \source{1} and a ``target'' \target into four nonnegative components.
The components identify information that is redundant, uniquely associated with \source{0}, uniquely associated with \source{1}, and synergistic:
\begin{align*}
  \I{\source{0}\source{1} : \target} =
    & \Ipart{\source{0}\sep\source{1}}
	  & \textrm{\emph{redundant}} \\
    & + \Ipart[\target \setminus \source{1}]{\source{0}}
	  & \textrm{\emph{unique from \source{0}}} \\
    & + \Ipart[\target \setminus \source{0}]{\source{1}}
	  & \textrm{\emph{unique from \source{1}}} \\
    & + \Ipart{\source{0}\source{1}} ~.
	  & \textrm{\emph{synergistic}}
\end{align*}
Furthermore, the mutual information between \source{0} and \target is decomposed into two components:
\begin{align*}
  \I{\source{0} : \target} =
    & \Ipart{\source{0}\sep\source{1}}
	  & \textrm{\emph{redundant}} \\
    & + \Ipart[\target \setminus \source{1}]{\source{0}}
    ~.
	  & \textrm{\emph{unique from \source{0}}}
\end{align*}
And, similarly:
\begin{align*}
  \I{\source{1} : \target} =
    & \Ipart{\source{0}\sep\source{1}}
    & \textrm{\emph{redundant}} \\
    & + \Ipart[\target \setminus \source{0}]{\source{1}}
	  ~.
	  & \textrm{\emph{unique from \source{1}}}
\end{align*}
In this way, PID relates the four component informations.
However, it does not uniquely determine how to quantify them.
To do this, a definition must be supplied for one of them and then the others follow.

This allows for a range of choices.
In the case that one wishes to directly quantify the unique informations \Ipart[\target \setminus \source{1}]{\source{0}} and \Ipart[\target \setminus \source{0}]{\source{1}}, a consistency relation must hold when they are computed independently:
\begin{align}
  \Ipart[\target \setminus \source{1}]{\source{0}} & + \I{\source{1} : \target} \nonumber \\
  & = \Ipart[\target \setminus \source{0}]{\source{1}} + \I{\source{0} : \target}
  ~.
  \label{eq:consistency}
\end{align}

\section{The Camel and the Elephant}
\label{sec:intuitions}

There are two common ways of thinking about PID.
These approaches differ only in the (implied) directionality of cause and effect---a property unspecified by PID.

In the first approach, one thinks of \source{0} and \source{1} as ``inputs'' that, when combined, produce \target, a ``output''.
While seemingly helpful labels, their use already imports an unwarranted semantics to the relationship between the three random variables.
In this, it inadvertently begs the main issue we wish to raise here, while at the same time illustrating the issue.

When taking this view of PID, one generally asks questions such as ``How much information in \source{0} is uniquely conveyed to \target?''.
From this vantage, considering the role of the individual channels $\source{0}
\to \target$ and $\source{1} \to \target$ might or might not help develop intuition.
Recalling the aphorism ``a camel is a horse designed by committee'', we call this the \emph{camel intuition} as particular input events \source{0} and \source{1} come together to describe an output \target.

In the second approach, one considers \source{0} and \source{1} as ``noisy observations'' or ``representations'' of a single underlying object \target.
When taking this view, one might ask a question such as ``How much information in \target is uniquely captured by \source{0}?''.
Under this, the individual channels $\target \to \source{0}$ and $\target \to \source{1}$ take on primary importance.
After the parable of the blind men describing an elephant, we call this the \emph{elephant intuition} since particular objects \target may be described by various, possibly partial, representations, \source{0} and \source{1}.

\section{The Pointwise Unique Distribution}
\label{sec:distribution}

The \emph{pointwise unique} distribution~\cite{finn2018pointwise} is given by the events and probabilities displayed in \cref{tab:dist}: at any time exactly one of \source{0} or \source{1} is a `1' or `2' and matches \target, while the other is `0'.
Let's now interpret this distribution by adopting the camel and elephant intuitions in turn.
We will see that they provide contradictory interpretations of the relationships between the variables.

\begin{table}
  \pntunqdist
  \caption{The \emph{pointwise unique} distribution.}
  \label{tab:dist}
\end{table}

Adopting the camel intuition, we consider the ways in which \source{0} influences \target.
It is easy to see that half of the time (\cref{tab:dist}'s \nth{1} and \nth{3} rows) \source{0} is unable to say anything about the state of \target.
The other half of the time (the \nth{2} and \nth{4} rows) \source{0} and \target are perfectly correlated, while \source{1} is ignorant as to their state.
Analogously, this is true when considering how \source{1} influences \target.
In this way, we interpret the distribution's PID as consisting entirely of unique informations.
The camel intuition is summarized in \cref{tab:intuitions}.

When adopting the elephant intuition, however, a strikingly different picture emerges.
Taking the viewpoint of \target, both single channel distributions $p(\source{0}|\target)$ and $p(\source{1}|\target)$ are identical.
So, any information shared with one must be redundantly shared with the other.
These channels do not allow one to determine the states of either \source{0} or \source{1}.
What is learned, however, is that exactly one of them matches \target, while the other is `0'.
Furthermore, removing the remaining uncertainty in the values of \source{0} and \source{1} requires observing one of them---a synergistic effect.
The resulting elephant analysis is also summarized in \cref{tab:intuitions}.

\begin{table}
  \begin{tabular}{lrr}
    \multicolumn{3}{c}{Decompositions by Intuition}                                    \\
    \toprule
                                                     & camel          & elephant       \\
    \midrule
    \Ipart{\source{0}\sep\source{1}}                 & \SI{0}{\bit}   & \SI{1/2}{\bit} \\
    \Ipart[\target \setminus \source{1}]{\source{0}} & \SI{1/2}{\bit} & \SI{0}{\bit}   \\
    \Ipart[\target \setminus \source{0}]{\source{1}} & \SI{1/2}{\bit} & \SI{0}{\bit}   \\
    \Ipart{\source{0}\source{1}}                     & \SI{0}{\bit}   & \SI{1/2}{\bit} \\
    \bottomrule
  \end{tabular}
  \caption{
    Camel and elephant intuitions applied to \cref{tab:dist}'s pointwise unique distribution.
    The camel intuition takes the view that \source{0} and \source{1} supply \target with unique informations, though only one of them at a time.
    The elephant intuition takes the view that \target provides both \source{0} and \source{1} with the same information, but it gets erased on the way to exactly one of them.
  }
\label{tab:intuitions}
\end{table}

In short, the two directional PID interpretations lead to contradictory quantifications.
From the viewpoint of camels, elephant approaches create redundancy where there is none.
From the vantage of elephants, camels draw distinctions where none exist.
This has been discussed by Ref.~\cite{ince2017measuring} regarding whether or not unique information should depend on \I{\source{0} : \source{1}}.
From the camel's point of view, ignoring this as a constraint may ``artificially correlate'' \source{0} and \source{1} and thereby inflate redundancy.
This viewpoint can be more directly illustrated by considering the intermediate distribution from which \Ibroja[]{}~\cite{bertschinger2014quantifying}---an elephant---computes unique information for the pointwise unique distribution:
\begin{table}[h!]
  \begin{tabular}{cccc}
    \toprule
    $\source{0}$ & $\source{1}$ & $\target$ & $\Pr$ \\
    \midrule
    0 & 0 & 1 & \nicefrac{1}{4} \\
    0 & 0 & 2 & \nicefrac{1}{4} \\
    1 & 1 & 1 & \nicefrac{1}{4} \\
    2 & 2 & 2 & \nicefrac{1}{4} \\
    \bottomrule
  \end{tabular}
\end{table}\\
From the elephant's view, \I{\source{0} : \source{1}} is irrelevant.

\section{Secret Key Agreement}
\label{sec:secrets}

\emph{Secret key agreement} is a fundamental concept within information-theoretic cryptography~\cite{maurer1993secret}.
The central idea is that if three parties, Alice, Bob, and Eve, observe some joint probability distribution $ABE \sim p(a, b, e)$ where Alice has access only to $a$, Bob $b$, and Eve $e$, is it possible for Alice and Bob to agree upon a secret key of which Eve has no knowledge.
The degree to which they may generate such a secret key immediately depends upon the structure of the joint distribution $ABE$.
It also depends upon whether Alice and Bob are allowed to publicly communicate.

Concretely, consider Alice, Bob, and Eve each receiving $n$ independent, identically distributed samples from $ABE$---Alice receiving $A^n$, Bob $B^n$, and Eve $E^n$.
A \emph{secret key agreement scheme} consists of functions $f$ and $g$, as well as a protocol for public communication ($h$) allowing either Alice, Bob, neither, or both to communicate.
In the case of a single party being permitted to communicate---say, Alice---she constructs $C = h(A^n)$ and then broadcasts it to all parties.
In the case that both parties are permitted communication, they take turns constructing and broadcasting messages of the form $C_i = h_i(A^n, C_{[0 \ldots i-1]})$ (Alice) and $C_i = h_i(B^n, C_{[0 \ldots i-1]})$ (Bob)~\cite{gohari2017coding}.

Formally, a secret key agreement scheme is considered $R$-achievable if for all $\epsilon > 0$:
\begin{align*}
  K_A &\stackrel{(1)}{=} f(A^n, C) \\
  K_B &\stackrel{(2)}{=} g(B^n, C) \\
  p(K_A = K_B = K) &\stackrel{(3)}{\geq} 1 - \epsilon \\
  \I{K : C E^n} &\stackrel{(4)}{\leq} \epsilon \\
  \frac{1}{n} \H{K} &\stackrel{(5)}{\geq} R - \epsilon
\end{align*}
where $(1)$ and $(2)$ denote the method by which Alice and Bob construct their keys $K_A$ and $K_B$, respectively, $(3)$ states that their keys must agree with arbitrarily high probability, $(4)$ states that the information about the key which Eve---armed with both her private information $E^n$ as well as the public communication $C$---be arbitrarily small, and $(5)$ states that the key consists of approximately $R$ bits per sample.

The greatest rate $R$ such that an achievable scheme exists is known as the \emph{secret key agreement rate}.
Notational variations indicate which parties are permitted to communicate.
In the case that Alice and Bob are not allowed to communicate, their rate of secret key agreement is denoted \SKARzero{A}{B}{E}.
When only Alice is allowed to communicate their secret key agreement rate is \SKARone{A}{B}{E}.
And, similarly, if only Bob is permitted to communicate.
When both Alice and Bob are allowed to communicate, their secret key agreement rate is denoted \SKARtwo{A}{B}{E}.
In this, we modified the standard notation for secret key agreement rates to emphasize which party or parties communicate.

In the case of no communication, \SKARzero{A}{B}{E} is given by~\cite{chitambar2018conditional}:
\begin{align}
  \SKARzero{A}{B}{E} = \H{A \meet B | E}
  \label{eq:skarzero}
\end{align}
where $X \meet Y$ denotes the G{\'a}cs-K{\"o}rner common random variable~\cite{gacs1973common}.
It is worth noting that this quantity does not vary continuously with the distribution and generically vanishes.

In the case of one-way communication, \SKARone{A}{B}{E} is given by~\cite{ahlswede1993common}:
\begin{align}
  \SKARone{A}{B}{E} = \max \left\{ \I{B : K | C} - \I{E : K | C} \right\}
  \label{eq:skarone}
\end{align}
where the maximum is taken over all variables $C$ and $K$, such that the following Markov condition holds: $C \markov K \markov A \markov BE$.
It suffices to consider $K$ and $C$ such that $|K| \leq |A|$ and $|C| \leq |A|^2$.

There are no such solutions for \SKARtwo{A}{B}{E}, however both upper- and lower-bounds are known~\cite{gohari2017coding}.

Let us now consider the pointwise unique distribution of \cref{tab:dist} and the ability of \source{0} and \target to agree upon a secret key while \source{1} eavesdrops.
\footnote{
  Secret key agreement rates have been associated with unique informations before.
  An upper bound on \SKARtwo{A}{B}{E}---the intrinsic mutual information~\cite{maurer1999unconditionally}---is known to not satisfy the consistency condition \cref{eq:consistency}~\cite{bertschinger2013shared}.
  More recently, the relationship between a particular method of quantifying unique information and one-way secret key agreement has been considered~\cite{banerjee2018unique}.
}
This can be interpreted four different ways.
First, neither \source{0} nor \target may be allowed to communicate.
Second, only \target can communicate.
Third, only \source{0} is permitted to communicate.
Finally, both \source{0} and \target may be allowed to communicate.
Note that the eavesdropper \source{1} is not allowed to communicate in any secret sharing schemes here.
Looking at this distribution, a general strategy becomes clear: both \source{0} and \target need some scheme to determine when they agree (the \nth{2} and \nth{4} rows).

Broadly, the only way in which both \source{0} and \target can come to understand if they match or not is if \source{0} is permitted to broadcast whether she observed a 0 or not.
Therefore, in the instances where \source{0} is not communicating there is no ability to agree upon a key: $\SKARzero{\source{0}}{\target}{\source{1}} = \SKARone{\target}{\source{0}}{\source{1}} = \SI{0}{\bit}$.
However, when \source{0} is allowed communication a key can be agreed upon: $\SKARone{\source{0}}{\target}{\source{1}} = \SKARtwo{\source{0}}{\target}{\source{1}} = \SI{1/2}{\bit}$.
\footnote{
  It is known that $\SKARtwo{\source{0}}{\target}{\source{1}} = \SI{1/2}{\bit}$ due to the convergence of upper and lower bounds in this instance.
}
These rates are summarized in \cref{tab:skars}.

\begin{table}
  \begin{tabular}{lr}
    \multicolumn{2}{c}{Secret Key Agreement Rates}              \\
    \toprule
    \SKARzero{\source{0}}{\target}{\source{1}} & \SI{0}{\bit}   \\
    \SKARzero{\source{1}}{\target}{\source{0}} & \SI{0}{\bit}   \\
    \midrule
    \SKARone{\target}{\source{0}}{\source{1}}  & \SI{0}{\bit}   \\
    \SKARone{\target}{\source{1}}{\source{0}}  & \SI{0}{\bit}   \\
    \midrule
    \SKARone{\source{0}}{\target}{\source{1}}  & \SI{1/2}{\bit} \\
    \SKARone{\source{1}}{\target}{\source{0}}  & \SI{1/2}{\bit} \\
    \midrule
    \SKARtwo{\source{0}}{\target}{\source{1}}  & \SI{1/2}{\bit} \\
    \SKARtwo{\source{1}}{\target}{\source{0}}  & \SI{1/2}{\bit} \\
    \bottomrule
  \end{tabular}
\caption{The variety of secret sharing schemes and their rates for the
  pointwise unique distribution of \cref{tab:dist}.
  }
  \label{tab:skars}
\end{table}

\section{Directionality, Naturalness, and Consistency}
\label{sec:natural}

We are now in a position to integrate the two intuitions with the results of secret key agreement rates.
The camel intuition, with the channels $\source{0} \to \target$ and $\source{1} \to \target$ taking center stage, most closely aligns with the one-way secret key agreement rates \SKARone{\source{0}}{\target}{\source{1}} and \SKARone{\source{1}}{\target}{\source{0}}.
This also agrees with \cref{sec:distribution}'s quantification (compare \cref{tab:intuitions,tab:skars}):
\begin{align*}
  \Ipart[\target \setminus \source{1}]{\source{0}} &=
  \SKARone{\source{0}}{\target}{\source{1}} ~\text{and} \\
  \Ipart[\target \setminus \source{0}]{\source{1}} &= \SKARone{\source{1}}{\target}{\source{0}}
  ~.
\end{align*}

The elephant intuition, with its focus on the channels $\target \to \source{0}$ and $\target \to \source{1}$ is more naturally aligned with the one-way secret key agreement rates \SKARone{\target}{\source{0}}{\source{1}} and \SKARone{\target}{\source{0}}{\source{1}}.
This again accords with \cref{sec:distribution}'s quantification:
\begin{align*}
  \Ipart[\target \setminus \source{1}]{\source{0}} &=
  \SKARone{\target}{\source{0}}{\source{1}} ~\text{and}\\
  \Ipart[\target \setminus \source{0}]{\source{1}} &= \SKARone{\target}{\source{1}}{\source{0}}
  ~.
\end{align*}
There are, however, difficulties with these approaches.

The first difficulty concerns the camel intuition.
If the one-way secret key agreement rates \SKARone{\source{0}}{\target}{\source{1}} and \SKARone{\source{1}}{\target}{\source{0}} are used to quantify the unique informations \Ipart[\target \setminus \source{1}]{\source{0}} and \Ipart[\target \setminus \source{0}]{\source{1}}, respectively, the consistency relation given by \cref{eq:consistency} is not necessarily satisfied.
Importantly, though, if \SKARone{\target}{\source{0}}{\source{1}} and \SKARone{\target}{\source{1}}{\source{0}} are used, the resulting PID is always consistent.
One concludes that the elephant intuition is the more natural of the two when using one-way secret key agreement rates to quantify unique informations.

There is another difficulty. PID is defined to be agnostic to directionality.
Furthermore, only one of the myriad proposed PID axioms is contingent on any inherent directionality---the Blackwell Property~\cite{rauh2017extractable} and it is an elephant.
In this sense, neither the camel nor the elephant intuitions are consistent with PID.
Again relating to secret key agreement, this implies that unique informations should more closely align with either the pair \SKARzero{\source{0}}{\target}{\source{1}} and \SKARzero{\source{1}}{\target}{\source{0}} or with the pair \SKARtwo{\source{0}}{\target}{\source{1}} and \SKARtwo{\source{1}}{\target}{\source{0}}; neither of which adopt any sort of directionality.

Both approaches bring their own further difficulties.
On the one hand, the no-communication secret key agreement rate is not continuous in the space of distributions, whereas PID is generally considered to vary continuously.
On the other hand, the two-way secret key agreement rate \SKARtwo{\source{0}}{\target}{\source{1}} has no known closed-form solution, only upper and lower bounds, and so it cannot be practically computed.
Furthermore and perhaps more fundamentally, whether or not the two-way secret key agreement rate results in a consistent decomposition is not known.
That said, our extensive searches of examples for which the upper and lower bounds converge are encouraging---they have not resulted in any violations of \cref{eq:consistency}.

\vspace{-0.1in}
\section{Conclusion}
\label{sec:conclusion}
\vspace{-0.1in}

At present, a primary barrier for PID's general adoption as a useful and possibly a central tool in analyzing how complex systems store and process information is an agreement on a method to quantify its component informations.
Here, we posited that one reason for disagreement stems from conflicting intuitions regarding the decomposition's operational behavior.
This suggests several possibilities.

The first is that PID is inherently context-dependent and quantification depends on a notion of directionality.
In this case, the elephant intuition is apparently more natural, as adopting closely related notions from cryptography results in a consistent PID.
If context demands the camel intuition, though, either a noncryptographic method of quantifying unique information is needed or consistency must be enforced by augmenting the secret key agreement rate.

The second possibility suggested by our observations is that intuitions which
project a directionality on the decomposition are inherently flawed and that any correct quantification must be independent of direction.
Interestingly, cryptographic notions may still play a role here.
Though, since there is as yet no known way to compute the two-way secret key agreement rate, its application remains open.

A final possibility is that associating secret key agreement rates with unique information is fundamentally flawed and that, ultimately, PID quantifies unique information as something distinct from the ability to agree upon a secret key.

Given that one of the main factors driving PID's creation was the need for interpretability, ensuring that the intuitions brought to bear are consistent with the quantitative values is of the utmost importance.
We described three quantitative regimes, each corresponding to a specific directionality or the lack thereof.
While it is possible that each can play a distinct role in the understanding of complex systems, our hope is that a single method will emerge as the most useful and accepted approach to understanding the organization of information within a joint probability distribution.

\vspace{-0.2in}
\section*{Acknowledgments}
\label{sec:acknowledgments}
\vspace{-0.1in}

All calculations herein were performed using the \texttt{dit} Python package~\cite{dit}.
We thank P. Banerjee, E. Olbrich, and D. Feldspar for many helpful discussions.
As a faculty member, JPC thanks the Santa Fe Institute and the Telluride Science Research Center for their hospitality during visits.
This material is based upon work supported by, or in part by, Foundational Questions Institute grant FQXi-RFP-1609, the U.S. Army Research Laboratory and the U.S. Army Research Office under contracts W911NF-13-1-0390 and W911NF-13-1-0340 and grant W911NF-18-1-0028, and via Intel Corporation support of CSC as an Intel Parallel Computing Center.

\end{document}